# 0.71-Å resolution electron tomography enabled by deep learning aided information recovery

Chunyang Wang[1,//], Guanglei Ding[1,2,//], Yitong Liu[2,*], Huolin L. Xin[1,*]
1. Department of Physics and Astronomy, University of California, Irvine, CA 92697, USA
2. School of Information and Communication Engineering, Beijing University of Posts and Telecommunications, Beijing 100876, China
[//] These authors contributed equally
*Correspondence should be addressed to YL (liuyitong@bupt.edu.cn) and HLX (huolinx@uci.edu)

**Electron tomography, as an important 3D imaging method, offers a powerful method to probe the 3D structure of materials from the nano- to the atomic-scale. However, as a grant challenge, radiation intolerance of the nano-scale samples and the missing-wedge-induced information loss and artifacts greatly hindered us from obtaining 3D atomic structures with high fidelity. Here, for the first time, by combining generative adversarial models with state-of-the-art network architectures, we demonstrate the resolution of electron tomography can be improved to 0.71 Å which is the highest three-dimensional imaging resolution that has been reported thus far. We also show it is possible to recover the lost information and remove artifacts in the reconstructed tomograms by only acquiring data from − 50 to + 50 degrees (44% reduction of dosage compared to -90 to +90 degrees full tilt series). In contrast to conventional methods, the deep learning model shows outstanding performance for both macroscopic objects and atomic features solving the long-standing dosage and missing-wedge problems in electron tomography. Our work provides important guidance for the application of machine learning methods to tomographic imaging and sheds light on its applications in other 3D imaging techniques.**

As the archetectural complexity of integrated circuits and functional nanomaterials advances into the atomic region, owing to its high resolution transmission electron microscopy (TEM) has become an essential tool in validating the structure of nanomaterials and devices. In contrast to conventional TEM imaging which only provides projected two-dimensional (2D) information, electron tomography is a powerful tool that can probe the 3D internal structure and chemistry of materials at the nano- and the atomic scale. Due to its high resolution, it has been widely used in biological [1, 2], physical[3] and materials sciences[4-6]. Recently, with the development of discrete tomography[7-9] and atomic-scale electron tomography[10-12], deciphering structure of materials atom by atom has become possible. Electron tomography renders 3D reconstruction of an object by taking a series of 2D projections from a wide range of orientations. Ideally, projection images from −90° to +90° around a fixed axis are needed to render a perfect three-dimensional (3D) reconstruction. However, due to the limited space in electron microscope or shadowing owing to the holder, it is practically impossible to obtain projections for the full tilt range. With specialized tomography holder, images from −70° to +70° can be obtained in most TEMs; whereas in liquid-cell tomography, the tilt range is very limited, typically < 45°, because the liquid flow holders are much bulkier [13]. The limited tilt range unavoidably lead to a 'missing wedge' of information which renders artifacts in the reconstructed tomograms. These artifacts introduced by the missing wedge reduce the resolution and reliability of the reconstructed tomograms and sometimes, can even lead to serious misinterpretations[4]. So far, the missing wedge problem has been the main source of systematic error that limits the application of electron tomography.

Mathematically speaking, when there are insufficient number of projections, the inverse problem of tomography is ill-posed because the solution is non-unique. Therefore, to constrain the solution space, strong priors need to be used to regularize the problem. For example, total variance minimization (TVM) combines iterative reconstruction and regularization of total variance to recover the lost information and reduced the artifacts introduced by the missing wedge[14, 15]. This method is inspired by compressive sensing [16, 17] and it essentially deploys the sparsity constraint in the gradient domain of the tomogram. One of the caveats of TVM is that it is not parameter free and computationally expensive. These aside, the real problem of TVM or any generalized TVM approach is that they are bound to one regulerizatoin that promotes one prior constraint on the solution which may or may not be suitable for the object of interest. For example, TVM promotes piecewise constants rendering cartoon-like tomogram lacking gradient details. Instead it is highly suitable to treat this problem as a classification and inpainting problem which involves the recognition of artifacts and the inpaiting of the 'correct' information.

Based on this consideration, for the first time, we introduced a deep-learning-based information-recovering model to recover the lost information and remove artifacts in the reconstructed tomograms of atomic electron tomography. Compared with the conventional methods, our deep learning model has free parameters and shows outstanding performance on both macroscopic and atomic objects. Particularly, we show that this method has improved the resolution of atomic electron tomography to 0.71 Å which is the highest resolution achieved by any 3D imaging method thus far.

## Result

**Training pipeline and model evaluation.** To train a deep learning model for information-recovery and artifact removal, a training library with a range of missing wedge and a collage of spatial features is built (Supplementary Table 1). To construct our datasets of the training database, three open data sources, MGH[18] and the national biomedical imaging archive (NBIA)[19] comprised of medical brain and tumor CT images, and ImageNet[20] labeled with dog, along with 'phantom faces' comprised of multiple random ovals and polygons with different intensity are included. All the datasets were reshaped and normalized to (256 x 256) with intensity ranging from 0 to 1, and then they are augmented as shown in Supplementary Table 2.

Next, based on the training library, we constructed a information recovery and de-artifact model (IRDM) by combining GAN and U-Net++[21]. Figure 1 shows the training pipeline of IRDM. In this model, U-Net++ is used as the generator for GAN. The generator keeps generating de-artifacted images to deceive the discriminator while the discriminator tries to judge whether the images are fake or ground truth. Then, the GAN loss for generator which is derived from the judgment of discriminator will contribute to the backward propagation to improve the generator. In the meantime, a GAN loss for discriminator will be calculated to strengthen the discriminator's capability. In addition, condition GAN is used to stabilize the training and thus improve the GAN's performance.

Based on a training dataset (D2 in Supplementary Table 1) we trained and evaluated the performance of IRDM model. Figure 2 shows the reconstructions of a random 'phantom face' using IRDM, WBP, SART and TVM respectively. With missing wedge ranging from 40° to 80°, the reconstruction by conventional WBP method shows severe information loss (blurring at top and bottom edge) and artifacts such as long tails and elongation along the vertical direction (See another example in Supplementary Fig. 1). Although the SART and TVM method can effectively suppress the long tail artifacts to some extent, however, the blur and elongation still exist. In contrast to SART and TVM, the tomograms processed by IRDM looks almost identical to the ground truth with no obvious long tails, blurs and elongation along the vertical direction, which indicates our deep learning model has outstanding information recovering and de-artifacting capability. With the missing wedge increased, both the information loss and artifacts become more prominent, however, the performance of IRDM still remains robust even with missing wedge up to 80°. Besides 'phantom face', real medical image, i.e., brain CT images are also used to further evaluate the effectiveness of IRDM. Fig. 3 shows the reconstruction of a brain CT image with IRDM, WBP, SART and TVM. In good agreement with the result in Fig. 2, IRDM also shows robust capability of information-recovering and deartifacting of the brain image with missing wedge up to 80°.

**Performance of IRDM model.** To test the influence of the composition of training features and the included range of missing wedges in the library on the performance of IRDM, datasets (see details in Supplementary Table 2) with different composition and ranges of missing wedge were created and tested. For D1-D3, the composition of the datasets are different while the training ranges of missing wedges are kept the same (40° to 60°). For datasets D3-D6, the composition of the datasets are exactly the same, while the training ranges of missing wedge are different. Fig. 2a shows the average peak-signal-to-noise ratio (PSNR) and structural similarity index measure (SSIM)[22] of SART, TVM and IRDM trained on datasets with different compositions but the same missing wedge from 40° to 60°. IRDM shows evidently improved PSNR compared with WBP, SART and TVM methods. SSIM also shows the same trend as PSNR. In particular, the IRDM trained on dataset D2 shows the highest PSNR hence the best performance. Fig. 2b shows the average NRMSE of TVM and IRDM trained on datasets with the same data composition but different ranges of missing wedge. Compared with dataset D1 which only comprises of 'phantom face' data, the more diverse dataset D2 which includes 'phantom faces' as well as tumor and brain data shows lower NRMSE, thereby improved information-recovery and de-artifact capability.

Moreover, by adding more diversity to the training dataset with ImageNet data labeled with dog, the IRDM3 model trained on D3 yeilds higher NRMSE, thereby reduced information-recovering and de-artifacting capability for data with missing wedge from 40° to 60°. Notably, however, for data with larger missing wedge beyond the training range, e.g., more than 65˚, IRDM trained on D3 shows much lower NRMSE than that trained on D2, which means although adding diversity to the training datasets may slightly degrade the performance of IRDM on data within the range of trained missing wedge angles, it will certianly improve the robustness of IRDM model in a larger range of missing wedge. Furthermore, we explored the influence of the training range of missing wedge on the performance of the IRDM. From D5➔D4➔D3➔D6, the training ranges of missing wedge are gradually widened. Figure 2c,d shows the NRMSE of IRDM model trained on D3-D6 and TVM method. The result shows that IRDM trained on dataset D5 with the narrowest training range of missing wedge shows lowest NRMSE, thereby best performance in range of missing wedge from 50° to 53°, however, at higher missing wedge beyond the training range (47.5°-52.5°), the performance of IRDM trained on D5 deteriorates rapidly. In contrast, for IRDM trained on D6 with the broadest training range of missing wedge (35°-65°), although its NRMSE is slightly higher than that trained on D5 and D4, it shows lowest NRMSE for data with missing wedge beyond 62.5°. Besides, it is noteworthy that IRDM trained on D6 always shows much lower NRMSE than that of conventional TVM methods.

**Application of IRDM in atomic electron tomography.** To confirm the validity and robustness of our deep learning model, for the first time, we applied IRDM to information-recovery and de-artifact for atomic electron tomography with a missing wedge of approximately 40°. Fig. 5a shows a representative view of the 3D atomic reconstruction of a gold crystal using conventional WBP method. The power spectrum (lower panel of Fig. 5a) in Fourier space shows Bragg spots corresponding to the lattice of the crystal (higher order Bragg spots represent higher resolution) and it is noteworthy that a region with information loss (missing wedge) evidently exist the reconstruction. By applying IRDM model to the WBP rconstruction, the artifacts introduced by missing wedge is effectively eliminated. More importantly, higher order Bragg spots as indicated by circles in Fig. 5b are recovered. It means the lost high frequency information in Fourier space caused by missing wedge can be effectively recovered and thus the resolution of atomic electron tomography can be significantly improved.

## Discussion and Conclusion

The performance of the deep learing model, i.e., IRDM is dependent on both the type and diversity of the training data. As shown in Fig. 2b, the performance of D1 which only contains 'phantom faces' is not as robust as D2 which contains three types of data including 'phantom faces, brain and tumor CT images', even though they have the same size of training data. Compared with D1, D2 is a more diverse dataset comprised of different types of data but still share some similarities. Therefore, the robustness of IRDM trained on D2 is much higher than that trained on D1. Moreover, by adding more diverse data, i.e., image of dogs, with distinct features from the data in D2, D3 shows a slightly worsened performance in the range of lower missing wedge, however, it shows higher robustness in the range of higher missing wedge. It indicates that, in general, the robustness and stability of IRDM model will be improved in a more broad range of missing wedge, if trained on a dataset with more diversity.

With large increase of missing wedge, the feature of the objects will be significantly changed by artifacts. For example, the reconstruction of the brain with the missing wedge of 40° shows a rounded shape similar to the original one. In contrast, with the missing wedge increased to 80°, the bottom of the brain turns into a sharp ridge which significantly deviates from the original shape of the brain. Different from Gaussian noise, the distribution of this kind of artifacts introduced by missing wedge will remarkably change with missing wedge angles, it can not be effectively eliminated by appling the IRDM model trained on a dataset with missing wedge far different from that of the data. Therefore, to get better performance of IRDM, training data sets with specific range of missing wedge angles in line with that of the data should be applied.

To make our deep learning model generally applicable to various objects with different features, the training database should be as large and diverse as possible, which will comsume massive computing resources and data sources. In practice, if the database is not big and diverse enough, the optimal strategy to get the best information-recovery and de-artifact effect is to try to find a minimum inclusion matrix (training dataset) which shares the highest feature similarity with the data. For example, for 'phantom faces' with missing wedge between 40° and 60°, training dataset (a) as indicated in Fig. 6 is the most appropriate training dataset. If we expect the model works for both brain images and 'phatom faces', training dataset (b) (for missing wedge between 40° and 60°) and (c) (for missing wedge around 80°) are appropriate choices. Data size is another factor which has influence on the performance of IRDM. For example, if the size of training dataset (b) is limited, the performance of IRDM on brain image or 'phantom face' data will be degraded. Then, as an alternative, a more diverse training dataset, e.g. dataset (d) with dog images added in could be used to improve the robustness of IRDM.

In this paper, we demonstrate a deep learning-based inpainting method enables us to achieve a record-high spatial resolution, 0.71 Å, for 3D imaging. The 0.71-Å resolution will significantly improve our capabilities to localize and differentiate atoms with different atomic numbers and local coordination environment. It opens the possibility to measure materials or even biological specimens atom by atom. This development thus has important applications in biological, physical and materials sciences.

## Methods

**Structure of generator and discriminator.** The IRDM is trained by GAN structure with U-Net++ as generator. U-Net++ is a encoding-decoding structure with cross-layer connections to compensate the feature information loss during the encoding-decoding process. The cross-layers between encoder and decoder are DenseNet blocks[23], which can enhance the flow of information and the efficiency of parameters utilization. Each Conv Layer in Supplementary Fig. 2 contains two alternating stacks of convolution layer, batch normalization, and SELU activation function[24]. A max pooling is used to reduce the dimension which is similar to an encoder. After max pooling for four times, we use Conv Layer with upsampling as an decoder to recover the dimension. The input of U-Net++ is a WBP iradon transform from a sinogram with missing wedge and the output is an image after information recovering and de-artifacting. For discriminator, as shown in Supplementary Fig. 3, we use a normal stack convolution layer structure similar to VGG block[25]. We stack convolution layer, batch normalization[26], LeakyReLU activation function and max pooling as a Conv Layer in the first four layers. For the fifth layer, we replace max pooling with average pooling to reduce information loss. A single convolution layer is used as the last output layer without activation function.

**Loss function.** In our model, different from the joint loss in standard U-Net++, we only use the output from right layer to compute loss. Least squares function (LSGAN)[27] is used to compute the loss for generator and discriminator as least squares loss is stable and straightforward with lower computational cost, and it can eliminate gradients vanishing. Furthermore, by replacing the absolute discriminator loss with relativistic discriminator loss (RaGAN), we get RaLSGAN[28] loss functions shown as follows:

$$L_{\text{RaLSD}} = [D(x_{\text{real}})-E[D(x_{\text{fake}})]-1]^2 + [D(x_{\text{fake}})-E[D(x_{\text{real}})]+1]^2] \quad (1)$$

$$L_{\text{RaLSG}} = [D(x_{\text{real}})-E[D(x_{\text{fake}})]+1]^2 + [D(x_{\text{fake}})-E[D(x_{\text{real}})]-1]^2] \quad (2)$$

$$x_{\text{fake}} = G(x_{\text{input}}) \quad (3)$$

Equation (1) and (2) are the loss functions for the discriminator and generator, respectively. $x_{\text{input}}$ and $x_{\text{output}}$ are the input and output data of the generator and discriminator, respectively. $x_{\text{real}}$ is ground truth. E[·] is average operation of minibatch when loading from the dataset.

**Training strategy.** The details of our training strategy is summarized in Supplementary Table 3. The total training epochs are 60. The training frequency ratio of discriminative and generative is 1:1. We set the learning rate as 1e-4, and the rate decays at the 45 and 55 epochs by multipling 0.1. The training weight decay is 1e-4. We set minibatch size as 32 and use four Nvidia 1080TI GPUs.

## Author contributions

HLX conceived idea and led the research. All authors participated in building models and writing the manuscript.

## Acknowledgment

This work is supported by the University of California, Irvine.## Reference

1. **J. Frank**, *Electron tomography: methods for three-dimensional visualization of structures in the cell* (Springer, New York ; London, 2006), 2nd ed.

2. **D. J. De Rosier and A. Klug**, Nature, **217**, 130 (1968).

3. **P. A. Midgley and M. Weyland**, Ultramicroscopy, **96**, 3, 413-431 (2003).

4. **P. A. Midgley and R. E. Dunin-Borkowski**, Nature materials, **8**, 271 (2009).

5. **S. Bals, B. Goris, L. M. Liz-Marzan, and G. Van Tendeloo**, Angewandte Chemie, **53**, 40, 10600-10610 (2014).

6. **L. Han, Q. Meng, D. Wang, Y. Zhu, J. Wang, X. Du, E. A. Stach, and H. L. Xin**, Nature communications, **7**, 13335 (2016).

7. **K. J. Batenburg, S. Bals, J. Sijbers, C. Kubel, P. A. Midgley, J. C. Hernandez, U. Kaiser, E. R. Encina, E. A. Coronado, and G. Van Tendeloo**, Ultramicroscopy, **109**, 6, 730-740 (2009).

8. **S. Van Aert, K. J. Batenburg, M. D. Rossell, R. Erni, and G. Van Tendeloo**, Nature, **470**, 374 (2011).

9. **B. Goris, S. Bals, W. Van den Broek, E. Carbo-Argibay, S. Gomez-Grana, L. M. Liz-Marzan, and G. Van Tendeloo**, Nature materials, **11**, 11, 930-935 (2012).

10. **M. C. Scott, C. C. Chen, M. Mecklenburg, C. Zhu, R. Xu, P. Ercius, U. Dahmen, B. C. Regan, and J. Miao**, Nature, **483**, 7390, 444-447 (2012).

11. **R. Xu, C. C. Chen, L. Wu, M. C. Scott, W. Theis, C. Ophus, M. Bartels, Y. Yang, H. Ramezani-Dakhel, M. R. Sawaya, H. Heinz, L. D. Marks, P. Ercius, and J. Miao**, Nature materials, **14**, 11, 1099-1103 (2015).

12. **J. Zhou, Y. Yang, Y. Yang, D. S. Kim, A. Yuan, X. Tian, C. Ophus, F. Sun, A. K. Schmid, M. Nathanson, H. Heinz, Q. An, H. Zeng, P. Ercius, and J. Miao**, Nature, **570**, 7762, 500-503 (2019).

13. **W. J. Dearnaley, B. Schleupner, A. C. Varano, N. A. Alden, F. Gonzalez, M. A. Casasanta, B. E. Scharf, M. J. Dukes, and D. F. Kelly**, Nano letters,   (2019).

14. **A. Boudjelal, Z. Messali, A. Elmoataz, and B. Attallah**, Journal of medical imaging and radiation sciences, **48**, 4, 385-393 (2017).

15. **B. Goris, W. Van den Broek, K. J. Batenburg, H. H. Mezerji, and S. Bals**, Ultramicroscopy, **113**, 120-130 (2012).

16. **E. J. Candès**, "Compressive sampling", in *Proceedings of the international congress of mathematicians* (Madrid, Spain, 2006), p. 1433-1452.

17. **D. L. Donoho**, IEEE Transactions on information theory, **52**, 4, 1289-1306 (2006).

18. **Q. Fan, T. Witzel, A. Nummenmaa, K. R. Van Dijk, J. D. Van Horn, M. K. Drews, L. H. Somerville, M. A. Sheridan, R. M. Santillana, and J. Snyder**, Neuroimage, **124**, 1108-1114 (2016).

19. <https://imaging.nci.nih.gov/nbia-search-cover/>.

20. **J. Deng, W. Dong, R. Socher, L.-J. Li, K. Li, and L. Fei-Fei**, "Imagenet: A large-scale hierarchical image database", in *2009 IEEE conference on computer vision and pattern recognition* (Ieee, 2009), p. 248-255.

21. **Z. Zhou, M. M. R. Siddiquee, N. Tajbakhsh, and J. Liang**, "Unet++: A nested u-net architecture for medical image segmentation", in *Deep Learning in Medical Image Analysis and Multimodal Learning for Clinical Decision Support* (Springer, 2018), p. 3-11.

22. **A. Hore and D. Ziou**, "Image quality metrics: PSNR vs. SSIM", in *2010 20th International Conference on Pattern Recognition* (IEEE, 2010), p. 2366-2369.
7

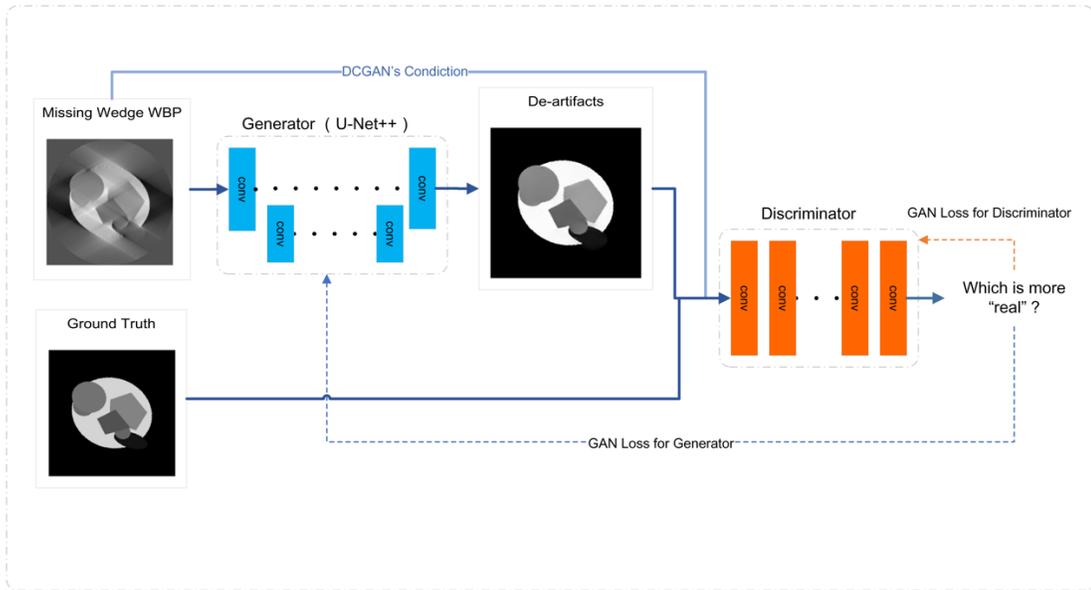

**Fig.1 | The training pipeline of the information-recovering and de-artifacting model (IRDM).**

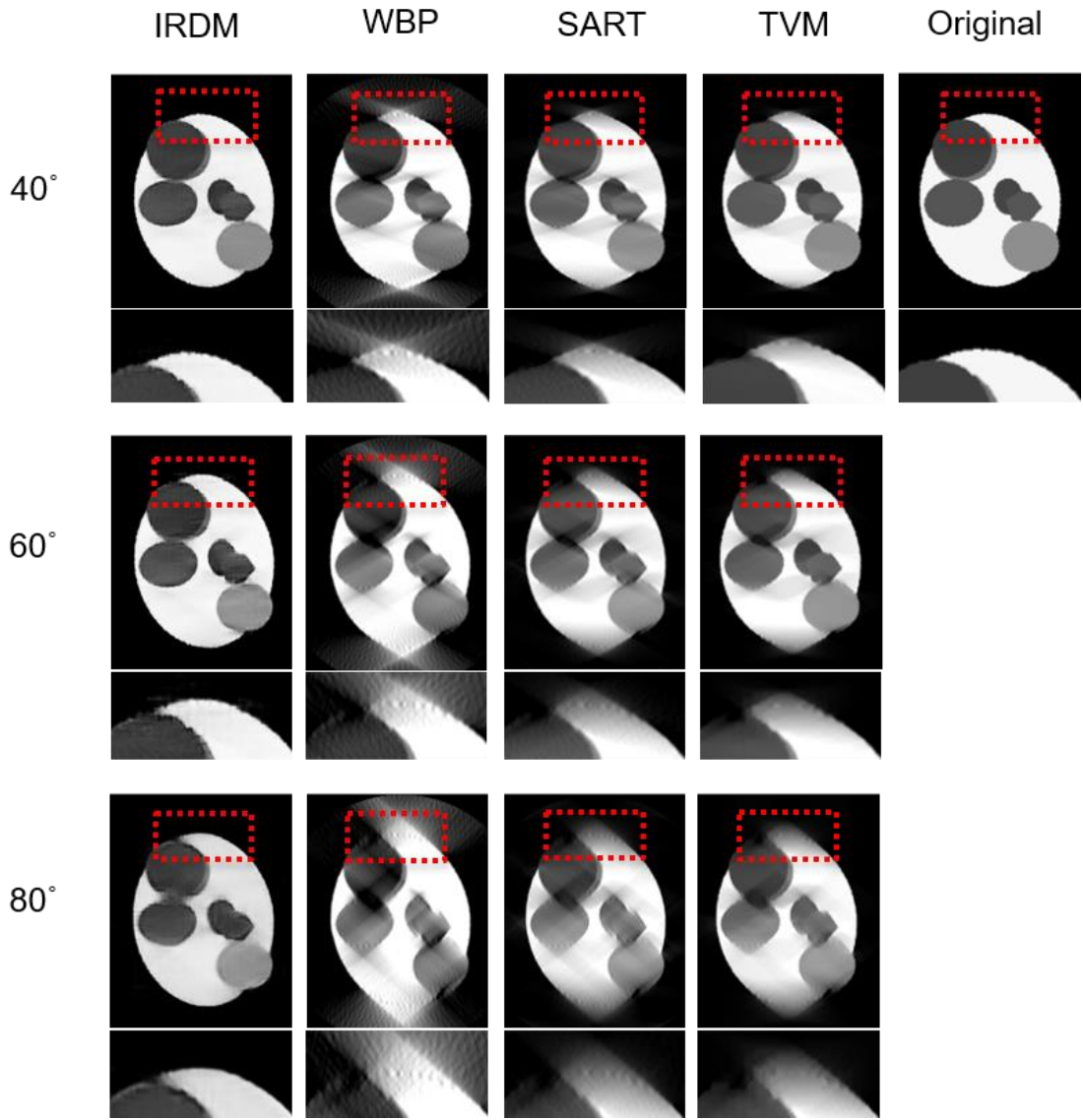

**Fig. 2 | Evaluation of IRDM in comparison with conventional methods by 'phantom face' reconstruction.** The 'phantom face' with different missing wedges are reconstructed by WBP, SART, TVM and IRDM, respectively. The lower panels show enlarged images of the boxed regions in upper panels.

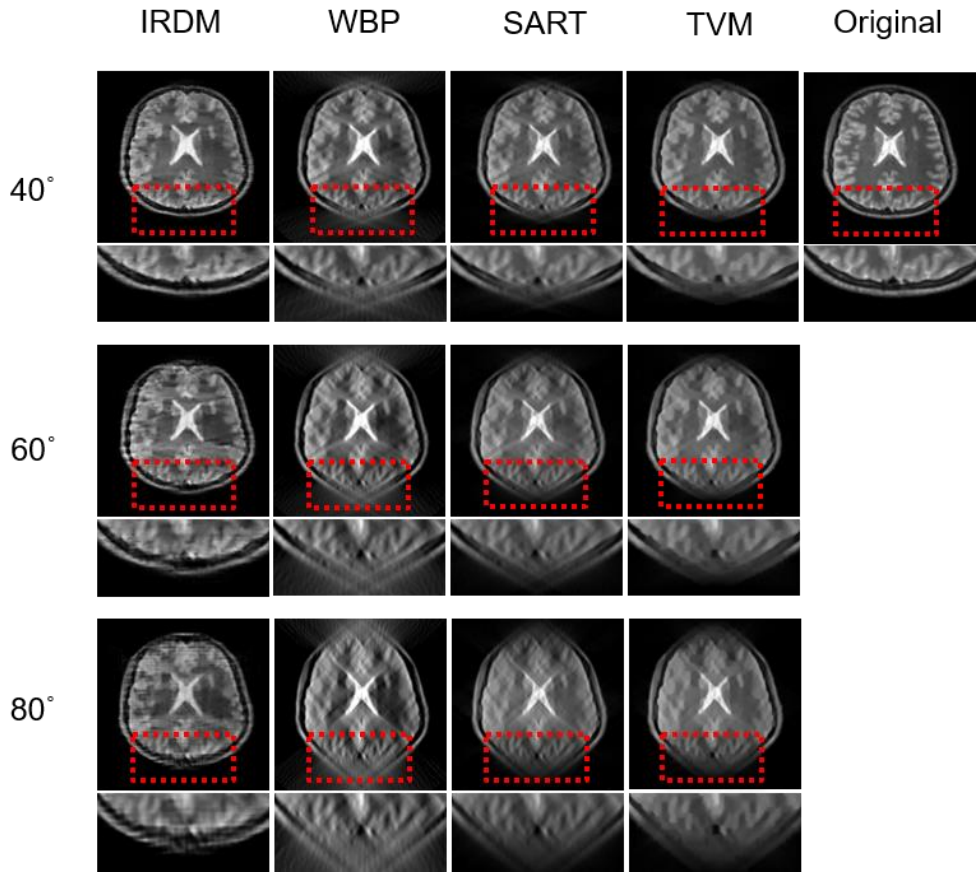

**Fig. 3 | Evaluation of IRDM in comparison with conventional methods by brain image reconstruction.** The brain image with different missing wedges are reconstructed by WBP, SART, TVM and IRDM, respectively. The lower panels show enlarged images of the boxed regions in upper panels.

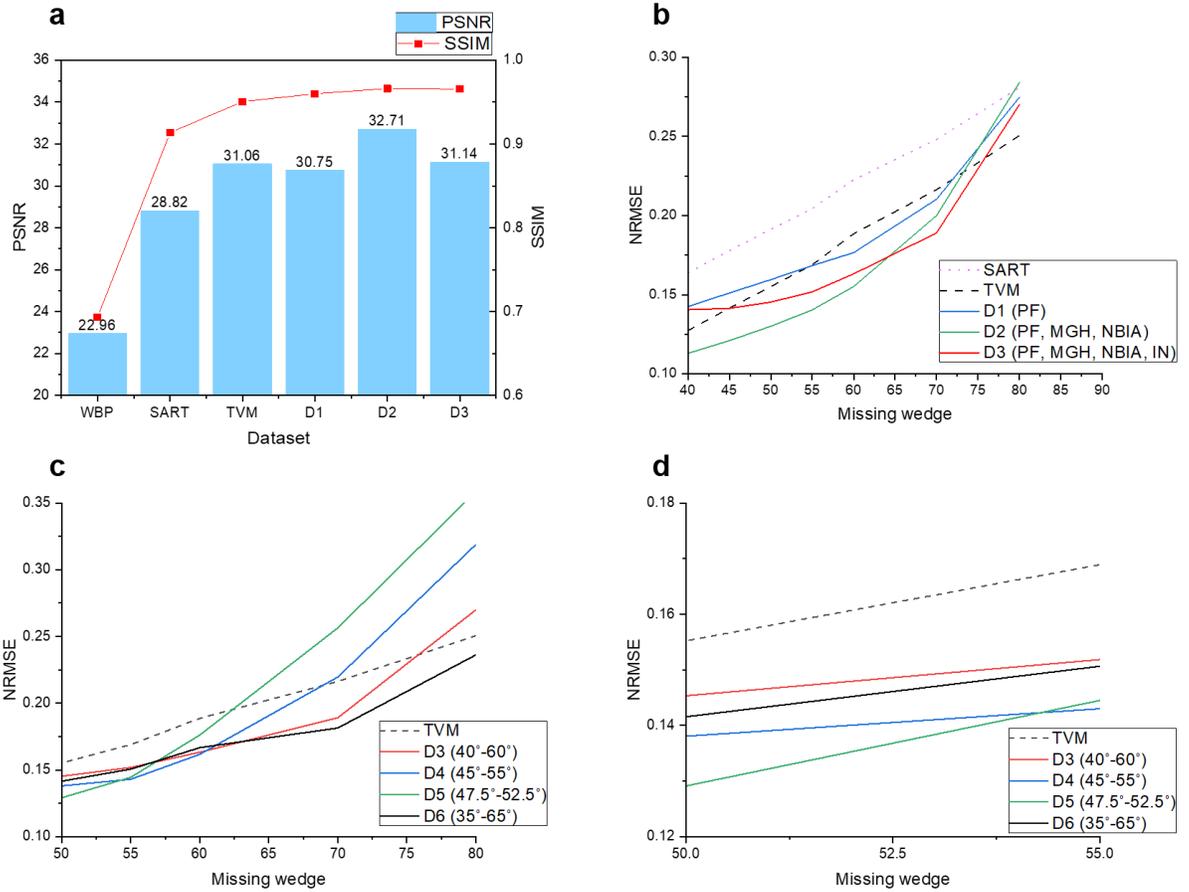

**Fig. 4 | The performance of IRDM, SART and TVM. a,** Average PSNR and SSIM of SART, TVM, and IRDM trained on datasets D1-D3. **b,** The NRMSE of SART, TVM and IRDM trained on D1-D3. **c,d,** The NRMSE of TVM and IRDM trained on D3-D6.

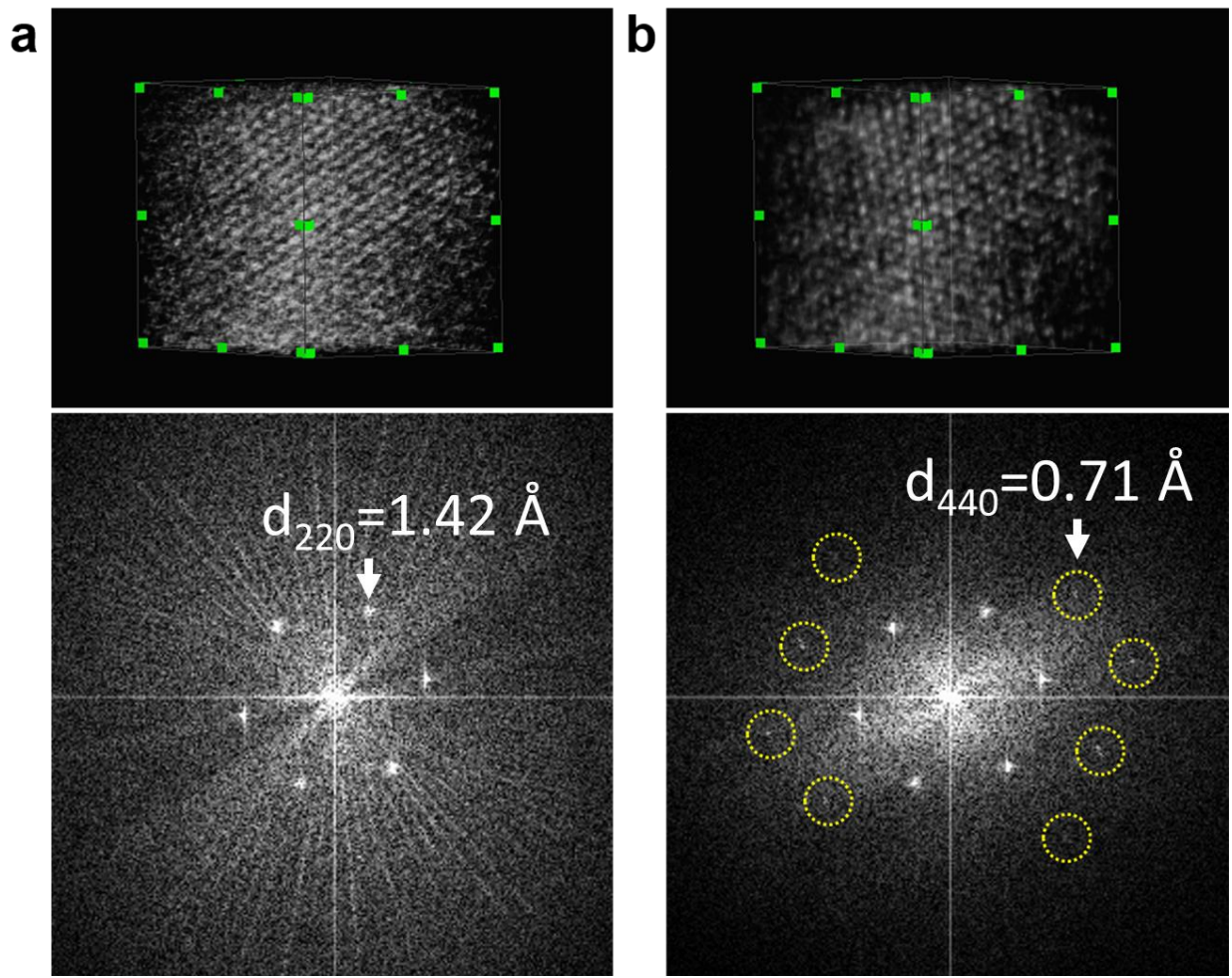

**Fig. 5 | 0.71-Å atomic electron tomography enabled by information inpainting. a,** 3D rendering (upper panel) of a gold crystal from 3D reconstruction by WBP and its corresponding power spectrum (lower panel) of the crystal. **b,** 3D rendering (upper panel) of the crystal in **a** processed by information recovery and de-artifact model (IRDM) and its corresponding power spectrum (lower panel). Yellow ovals indicate the recovered 440 Bragg spots by IRDM which corresponds to information transfer up to a spacing of 0.71 Å.

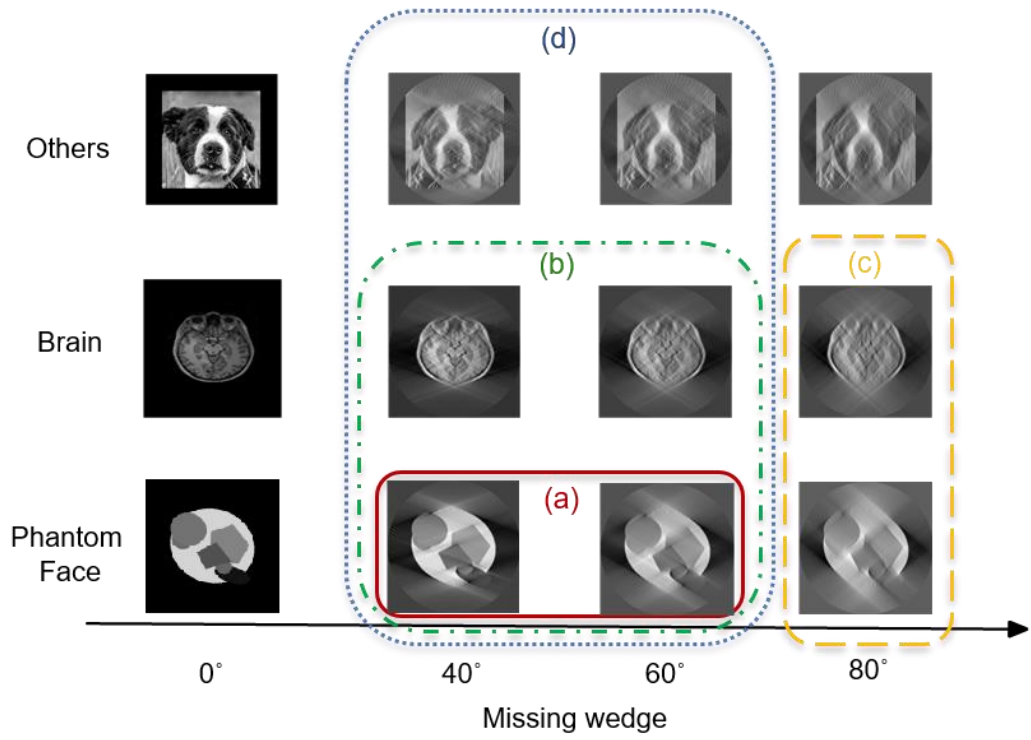

**Fig. 6** | Minimum inclusion matrix principle fro dataset selection for IRDM.

# Supplementary Materials for

# 0.71-Å resolution electron tomography enabled by deep learning aided information recovery

Supplementary Table 1: The constitutions of the training datasets.

| dataset | 'phantom face' | MGH | NBIA | ImageNet | Sum | Missing wedge angles(°) |
|---|---|---|---|---|---|---|
| Dataset 1(D1) | 45,000 | 0 | 0 | 0 | 45,000 | 40-60 |
| Dataset 2(D2) | 25,000 | 10,000 | 10,000 | 0 | 45,000 | 40-60 |
| Dataset 2'(D2') | 25,000 | 10,000 | 10,000 | 0 | 45,000 | 40-80 |
| Dataset 3(D3) | 25,000 | 10,000 | 10,000 | 10,000 | 55,000 | 40-60 |
| Dataset 4(D4) | 25,000 | 10,000 | 10,000 | 10,000 | 55,000 | 45-55 |
| Dataset 5(D5) | 25,000 | 10,000 | 10,000 | 10,000 | 55,000 | 47.5-52.5 |
| Dataset 6(D6) | 25,000 | 10,000 | 10,000 | 10,000 | 55,000 | 35-65 |

**Supplementary Table 2: The data augmentation for different data source.**

| datasources | Pad Resize | Radom Rotation | Radom Flip | Radom Affine |
|---|---|---|---|---|
| 'phantom face' | √ | √ | | |
| MGH | √ | √ | √ | √ |
| NBIA | √ | √ | √ | √ |
| ImageNet | √ | | | |

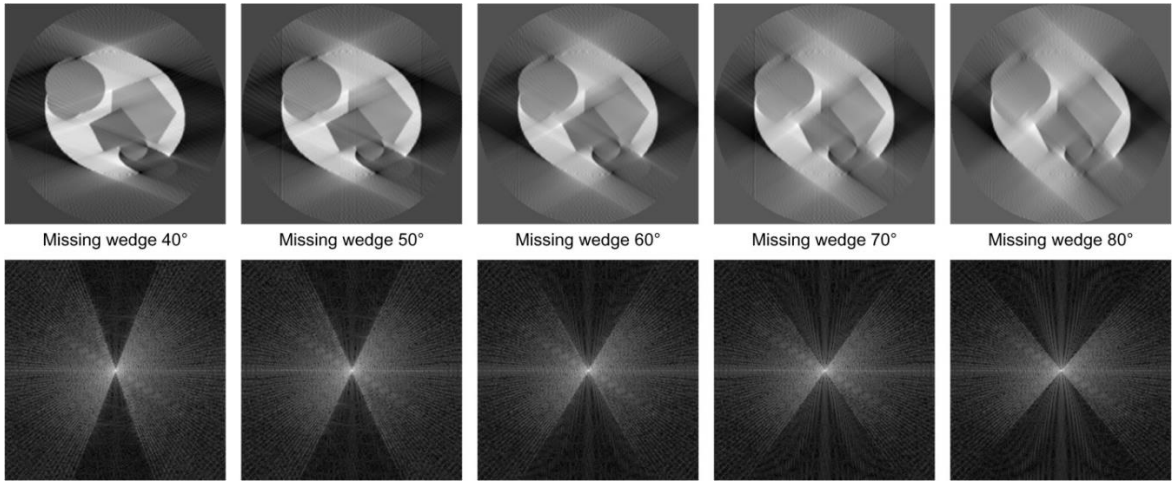

**Supplementary Figure 1: WBP reconstruction of 'phantom faces' with different missing wedge angles and their corresponding power spectra.**

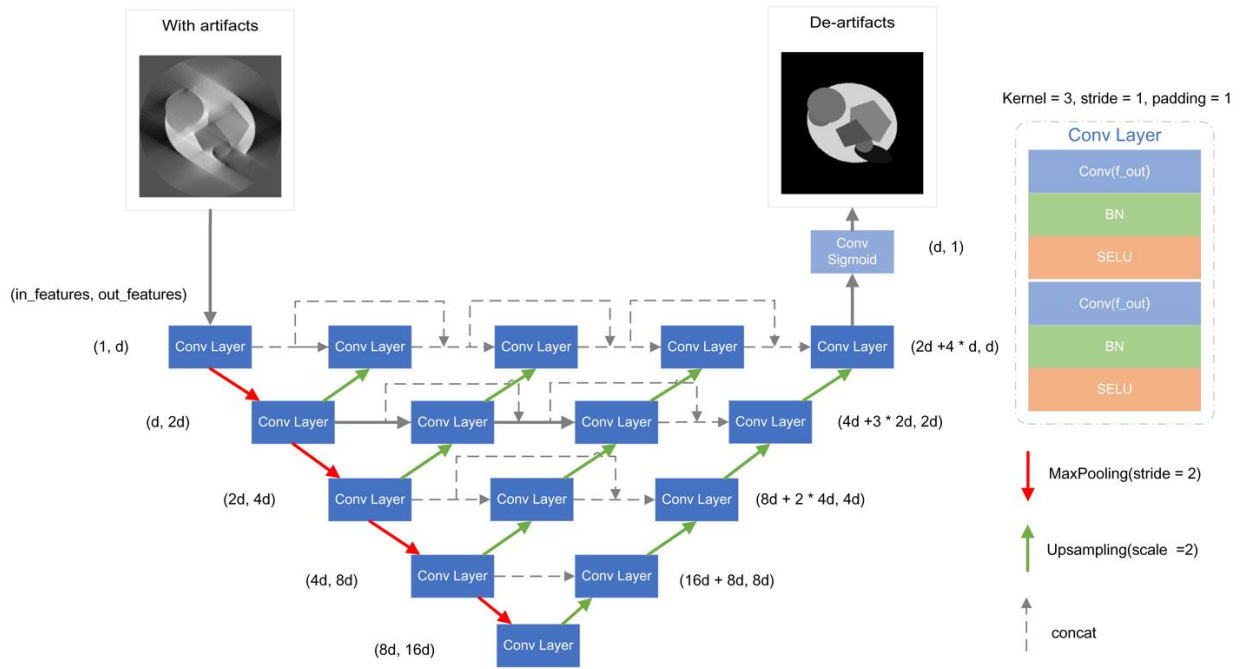

**Supplementary Figure 2: The Structure of U-net ++ model.**

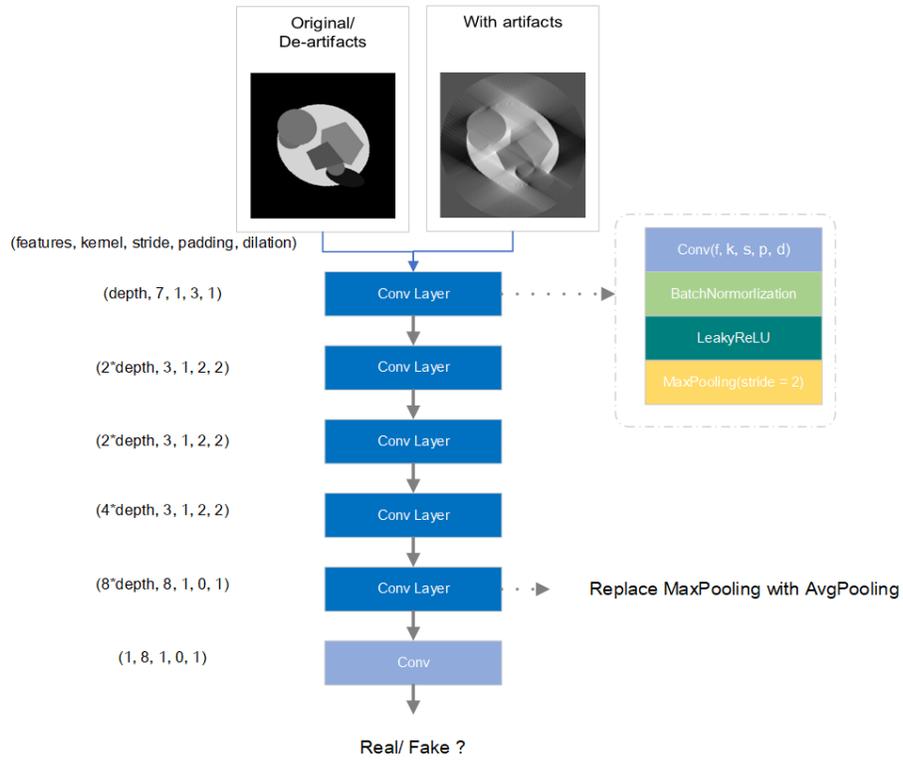

**Supplementary Figure 3: The structure of discriminator model.**

**Supplementary Table 3: The hyperparameters of optimizers for the generator and descriminator.**

| Optimizer | Learning rate | Weight decay | Alpha | Betas |
|---|---|---|---|---|
| Adam(generator) | 1e-4 | 1e-4 | \ | (0.9, 0.99) |
| RMSprop(discriminator) | 1e-4 | 1e-4 | 0.99 | \ |